%% file: main.tex
\documentclass[conference]{IEEEtran}
\IEEEoverridecommandlockouts

\usepackage{cite}
\usepackage{amsmath,amssymb,amsfonts}
\usepackage{algorithmic}
\usepackage{graphicx}
\usepackage{hyperref}
\usepackage{textcomp}
\usepackage{xcolor}
\usepackage{subcaption}
\usepackage{url}
\usepackage{cleveref}
\usepackage{comment}

\usepackage{subfiles}
\usepackage{threeparttable}

\def\BibTeX{{\rm B\kern-.05em{\sc i\kern-.025em b}\kern-.08em
    T\kern-.1667em\lower.7ex\hbox{E}\kern-.125emX}}
\begin{document}

\title{AI-based Resource Allocation: Reinforcement Learning for Adaptive Auto-scaling in Serverless Environments}

\author{
\IEEEauthorblockN{Lucia Schuler}
\IEEEauthorblockA{
\textit{Karlsruhe Institute of Technology}\\
\href{mailto:lucia.schuler@alumni.kit.edu}{lucia.schuler@alumni.kit.edu}
}
\and
\IEEEauthorblockN{Somaya Jamil}
\IEEEauthorblockA{
\textit{IBM Research \& Development GmbH}\\
\href{mailto:jamilsom@de.ibm.com}{jamilsom@de.ibm.com}
}
\and
\IEEEauthorblockN{Niklas K\"uhl}
\IEEEauthorblockA{
\textit{IBM Deutschland GmbH}\\
\textit{Karlsruhe Institute of Technology}\\
\href{mailto:niklas.kuehl@kit.edu}{niklas.kuehl@kit.edu}
}
}

\maketitle

\begin{abstract}
Serverless computing has emerged as a compelling new paradigm of cloud computing models in recent years. 
It promises the user services at large scale and low cost while eliminating the need for infrastructure management. 
On cloud provider side, flexible resource management is required to meet fluctuating demand. It can be enabled through automated provisioning and deprovisioning of resources.
A common approach among both commercial and open source serverless computing platforms is workload-based auto-scaling, where a designated algorithm scales instances according to the number of incoming requests.
In the recently evolving serverless framework Knative a request-based policy is proposed, where the algorithm scales resources by a configured maximum number of requests that can be processed in parallel per instance, the so-called concurrency.
As we show in a baseline experiment, this predefined concurrency level can strongly influence the performance of a serverless application.
However, identifying the concurrency configuration that yields the highest possible quality of service is a challenging task due to various factors, e.g. varying workload and complex infrastructure characteristics, influencing throughput and latency.
While there has been considerable research into intelligent techniques for optimizing auto-scaling for virtual machine provisioning, this topic has not yet been discussed in the area of serverless computing.
For this reason, we investigate the applicability of a reinforcement learning approach, which has been proven on dynamic virtual machine provisioning, to request-based auto-scaling in a serverless framework. 
Our results show that within a limited number of iterations our proposed model learns an effective scaling policy per workload, improving the performance compared to the default auto-scaling configuration.

\end{abstract}


\section{Introduction}
\subfile{sections/introduction}

\section{Background}
\subfile{sections/background}

\section{Related Work}
\subfile{sections/relatedwork}

\section{Experimental Setup}
\subfile{sections/experimentalsetup}

\section{Baseline Experiment}
\subfile{sections/baselineexperiment}

\section{Reinforcement Learning Experiment}
\subfile{sections/qlearningexperiment}


\section{Conclusion}
\subfile{sections/conclusion}

\bibliographystyle{./bibliography/IEEEtran}
\bibliography{./bibliography/references}

\end{document}

%% file: sections/introduction.tex
Driven by the advancements and proliferation of virtual machines (VMs) and container technologies, the adoption of serverless computing models has increased in recent years \cite{castro2019server}.
According to the Cloud Native Computing Foundation, serverless computing offers two main advantages to the user \cite{CNCFwhitepaper}.
First, with a true and fine-grained pay-as-you-go pricing model, costs only occur when resources are actually used and not for idle VMs or containers.
Second, there is no overhead for the user associated with infrastructure maintenance, such as provisioning, updating, and managing the server resources, as this is delegated to the cloud provider.
This also includes flexible on-the-fly scalability which enables resources to be added or removed automatically depending on the incoming load. 
For providers, the auto-scaling capability provides the ability to optimize resource utilization and reduce the effort required to manage cloud-scale applications \cite{castro2019server}.

In the implementation, the scaling mechanisms differ within the serverless offerings. 
Some open source serverless frameworks use the resource-based Kubernetes \textit{Horizontal Pod Autoscaler} (HPA) 
to drive scaling via per-instance CPU or memory utilization thresholds (e.g. 
Fission \cite{fission}).
This, of course, makes the auto-scaling feature dependent on the fast and correct calculations of respective system components \cite{li2019understanding}.  
Commercially provided serverless platforms often feature workload-based scaling by providing additional resources when incoming traffic increases, e.g. AWS Lambda initializes an instance for each new request coming in until a limit is reached \cite{AWSLambda}.
However, the creation of a new instance implies a certain time lag, known as \textit{cold start}.
To bypass this issue to a certain extent, a recently emerging open-source framework \textit{Knative} supports parallel processing of up to a predefined number of concurrent requests per instance \cite{knativedocsautoscaling}.
When the so-called \textit{concurrency} is reached, \textit{Knative Pod Autoscaler} (KPA) deploys additional pods to handle the load.
Moreover, the concurrency parameter can be adjusted manually to use resources more efficiently and to adapt the auto-scaling system to individual workloads. 

In the work at hand we show that, depending on the workload, different concurrency levels can influence the performance and can lead to a latency difference of up to multiple seconds.
Since this can have a critical impact on the user experience in serverless computing, we propose a reinforcement learning (RL) based model to dynamically determine the optimal concurrency for an individual workload.
In general, RL formalizes the idea of an agent learning effective decision-making policies through a sequence of trial-and-error interactions with its environment.
Thereby, the agent evaluates the current state of the system dynamics in each iteration, and then decides on a particular action.
After the action has been performed, the agent receives either positive or negative reward 
and consequently learns about the goodness of the respective action-state combination.
As this approach does not require any prior knowledge about incoming workload and can adapt to changes at runtime, RL algorithms have been proven as valid methods in the field of VM auto-scaling techniques in research \cite{lorido2014review}. 
However, it has not been studied in a serverless environment.
Therefore, we evaluate the applicability of the established RL-algorithm Q-learning to determine the concurrency level with optimized performance.

Specifically, we implement a cloud-based framework upon which two consecutive experiments are conducted.
First, we perform an extensive analysis to examine performance variations of different workload profiles under different auto-scaling configurations.
We demonstrate the dependence of throughput and latency on the concurrency level and indicate the potential for improvement through adaptive scaling settings.
Using these results, we enhance the framework with an intelligent RL-based logic to evaluate the ability of a self-learning algorithm for effective decision making in a serverless framework.
As we show in a second experiment, our proposed model is able to learn in limited time an appropriate scaling policy without prior knowledge of the incoming workload, resulting in an increased performance compared to the framework's default auto-scaling settings.

The remainder of the work is organized as follows.
Section \ref{sec:background} introduces the serverless platform Knative and the theory of Q-learning.
Section \ref{sec:relatedwork} reviews related work in both serverless frameworks and cloud-based auto-scaling techniques.
Section \ref{experimentalsetup} gives an overview of the underlying experimental setup of the work, based on which section \ref{sec:baseline} presents the tests on the impact of different concurrency limits.
Using these findings, section \ref{sec:qlearning} proposes a Q-learning model to adapt the concurrency limit on-the-fly.
Section \ref{conclusion} concludes the paper with remarks on limitations and possible future work.

%% file: sections/background.tex
 \label{sec:background}

To allow for a common understanding of the application domain and used techniques, we first provide an overview of the functionality of Knative and its auto-scaling feature.
Further, we introduce the theoretical foundations of the Q-learning algorithm which is applied in the second experiment.

\subsection{Knative Serverless Platform} \label{subsec:Knative}

As an open-source serverless platform, Knative provides a set of Kubernetes-based middleware components to support deploying and serving of serverless applications, including the capability to automatically scale resources on demand \cite{knativedocsautoscaling}.
%

\begin{figure}[t]
\centering
\includegraphics[width=8.5cm]{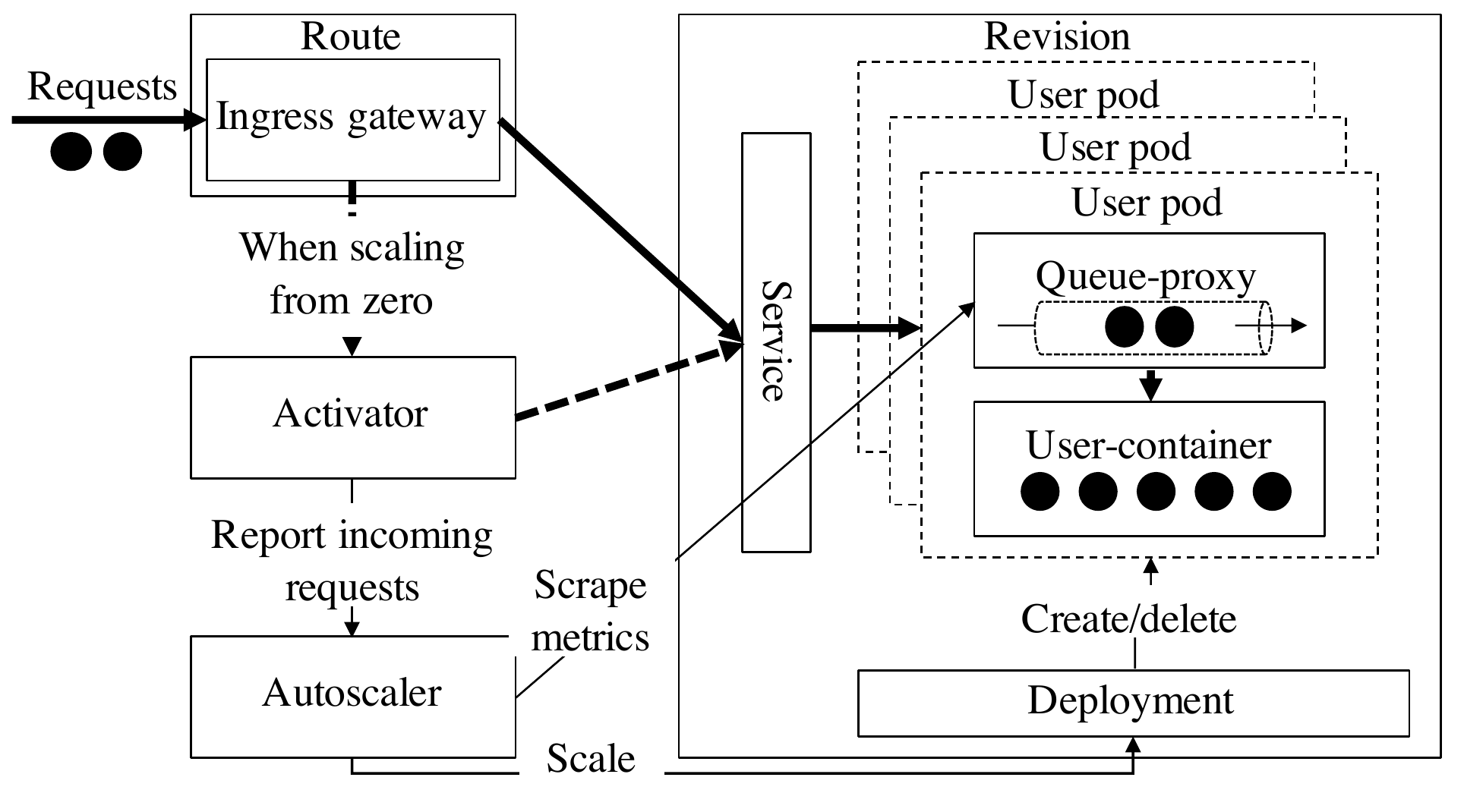}
\caption{Simplified request flow in Knative} 
\label{fig:knative-autoscaling}
\end{figure}

The auto-scaling function is implemented by different serving components, described by the request flow in \Cref{fig:knative-autoscaling} based on Knative v0.12.
If a service revision is scaled to zero, i.e. the service deployment is reduced to a replica of null operating pods, the ingress gateway forwards incoming requests first to the activator \cite{knativedocsautoscaling}.
The activator then reports the information to the autoscaler, which instructs the revision's deployment to scale-up appropriately.
Further, it buffers the requests until the user pods of the revision become available, which can cause cold-start costs in terms of latency, as the requests are blocked for the corresponding time.
In comparison, if a minimum of one replica is maintained active, the activator is bypassed and the traffic can flow directly to the user pod.

When the requests reach the pod, they are channeled by the \emph{queue-proxy} container and, subsequently, processed in the \emph{user-container}.
The queue-proxy only allows a certain number of requests to enter the user-container simultaneously, and queues the requests if necessary.
The amount of parallel processed requests is specified by the \textit{concurrency} parameter configured for a particular revision.
By default, the value is set to a concurrency target of 100, defining how many parallel requests are preferred per user-container at a given time.
However, the user can explicitly restrict the number of concurrent requests by specifying a value between 0 and 1000 for the \textit{concurrency limit}.\footnote{A value of 0 allows unlimited concurrent requests and therefore results in no scaling \cite{openshiftknative}.}
Further, each queue-proxy measures the incoming load and reports the average concurrency and requests per second on a separate port.
The metrics of all queue-proxy containers are scraped by the autoscaler component, which then decides how many new pods need to be added or removed to keep the desired concurrency level.

\subsection{Q-learning}

RL refers to a collection of trial-and-error methods in which an agent is trained to make good decisions by interacting with his environment and receiving positive or negative feedback in form of rewards for a respective action. 
A popular RL algorithm is the model-free Q-learning.

Q-learning stepwise trains an approximator $Q_{\theta}(s,a)$ of the optimal action-value function $Q^*$. 
$Q_{\theta}(s,a)$ specifies the cumulated reward the agent can expect when starting in a state $s$, taking an action $a$, and then acting according to the optimal policy forever after.
By observing the actual reward in each iteration, the optimization of the Q-function is performed incrementally per step $t$: 
\[Q(s_t,a_t)\xleftarrow{}(1-\alpha)Q(s_t,a_t) + \alpha[r_t + \gamma \max_{a} Q(s_{t+1},a)]\]
$\alpha$ describes the learning rate, i.e. to what extent newly observed information overrides old information and $\gamma$ a discount factor that serves to balance between the current and future reward.
As RL is a trial-and-error method, during training, the agent has to choose between the exploration of a new action and the exploitation of the current best option \cite{sutton2018rl}.
In research, this is often implemented with an $\epsilon$-greedy strategy, where $\epsilon$ defines the probability of exploration that usually decreases as the learning process advances \cite{rossi2019horizontal, mnih2013playing}.
With a probability of $1-\epsilon$, the agent selects based on the optimal policy and chooses the action that maximizes the expected return from starting in $s$, i.e. the action with the highest Q-value:
\[a^*(s) = \arg \max_{a} Q^*(s,a)\]
In the basic algorithm, the Q-values for each state-action combination are stored in a lookup table, the so-called \textit{Q-table}, indexed by states and actions. 
The tabular representation of the agent's knowledge serves as a basis for decision-making during the entire learning episode. 

%% file: sections/relatedwork.tex
 \label{sec:relatedwork}

To the best of our knowledge, the applicability of RL-based technology to optimize auto-scaling capabilities in serverless environments has not been investigated. However, considering the areas of serverless and intelligent auto-scaling separately, a large body of knowledge is available, summarized in the following subsections.

\subsection{Serverless computing}
With the growing number of serverless computing offerings, there has been an increasing interest of the academic community in comparing different solutions, with scalability being one of the key elements of evaluation \cite{li2019understanding}. 
In multiple works, different propriety serverless platforms were benchmarked, including their ability to scale, focusing on Amazon Lambda, Microsoft Azure Functions \cite{lloyd2018serverless}, along with Google Cloud Functions \cite{wang2018peeking} and additionally IBM Cloud Functions \cite{lee2018evaluation}.
Similar studies have been carried out in the area of open-source serverless frameworks, with greater attention paid to the auto-scaling capabilities.
Mohanty et al. \cite{mohanty2018evaluation} evaluated Fission, Kubeless, and OpenFaaS and concluded that Kubeless provides the most consistent performance in terms of response time. 
Another comparison of both qualitative and quantitative features of Kubeless, OpenFaas, Apache Openwhisk, and Knative, comes to the same conclusion, albeit generally indicating the limited user control over custom Quality of Service (QoS) requirements \cite{palade2019evaluation}.
These studies solely consider the default auto-scaler Kubernetes HPA.
Possible adjustments to the auto-scaling mechanism itself are not further examined.
Li et al. \cite{li2019understanding} propose a more concrete distinction between resource-based and workload-based scaling policies. 
The authors compare the performance of different workload scenarios using the tuning capability of concurrency levels in Knative and clearly suggest further investigation of the applicability of this auto-scaling capability, which also motivates this research.

\subsection{Auto-scaling}
As elasticity is one of the main characteristics of the increasing adaption of cloud computing, the automatic, on-demand provisioning and de-provisioning of cloud resources have been the subject of intensive research in recent years \cite{singh2019research}.
We discuss related work under two aspects: first, the underlying theories on which auto-scaling is built with a focus on RL, and second, the entities being scaled.

To classify numerous techniques at the algorithmic level, different taxonomies were proposed, where the predominant categories are threshold-based rules, queuing theory and RL \cite{lorido2014review,singh2019research}.
In the former, scaling decisions are made on predefined thresholds and are most popular among public cloud providers, e.g. Amazon ECS \cite{AWSservicescaling}.  
Despite the simplistic implementation, identifying suitable thresholds requires expert knowledge \cite{singh2019research}, or explicit application understanding \cite{dutreilh2010data}. 
Queuing theory has been used to mathematically model applications \cite{lorido2014review}. 
As they usually impose a stationary system, the models are 
less reactive towards changes \cite{lorido2014review}.

In contrast, RL offers an interesting approach through online learning of the most suitable scaling action and without the need for any a-priori knowledge \cite{lorido2014review}.
Many authors have therefore investigated the applicability of model-free RL algorithms, such as Q-learning, in recent years \cite{rossi2019horizontal}.
Dutreihl et al. \cite{dutreilh2010data} show that although Q-learning based VM controlling requires an extensive learning phase and adequate system integration, it can lead to significant performance improvements compared to threshold-based auto-scaling, since thresholds are often set too tightly while seeking for the optimal resource allocation.
To combine the advantages of both, Q-learning itself can be used to automatically adapt thresholds to a specific application \cite{horovitz2018efficient}.

In terms of the entity being scaled, RL has been mostly applied to policies for VM allocation and provisioning, e.g. in \cite{bitsakos2018derp}.
With the emergence of container-based applications, this field has become a greater focus of research \cite{rossi2019horizontal}. 
In both areas, the scope of action is concentrated mainly on horizontal (scale-out/-in) \cite{horovitz2018efficient}, vertical scaling (scale-up/-down) \cite{rao2011distributed}, or the combination of both \cite{rossi2019horizontal}.
However, little research has been done in areas that extend the classic auto-scaling problem of VM or container configuration.

As a novel approach we investigate the applicability of Q-learning to request-based auto-scaling in a serverless environment. 
Differently from the existing work on direct vertical or horizontal scaling with RL, we propose a model that learns an effective scaling policy by adapting the level of concurrent requests per container instance to a specific workload.

%% file: sections/experimentalsetup.tex
 \label{experimentalsetup}

To investigate different concurrency configurations, a flexible Kubernetes-based framework is designed which can be extended by an intelligent RL-based logic. 
In this section, we present the fundamental setup starting with the cloud architecture.
This section provides the foundation for both the first experiment assessing the impact of concurrency changes and the second experiment evaluating RL-based auto-scaling.

\subsection{Cloud Architecture}
\begin{figure}[t]
\centering
\includegraphics[width=8cm]{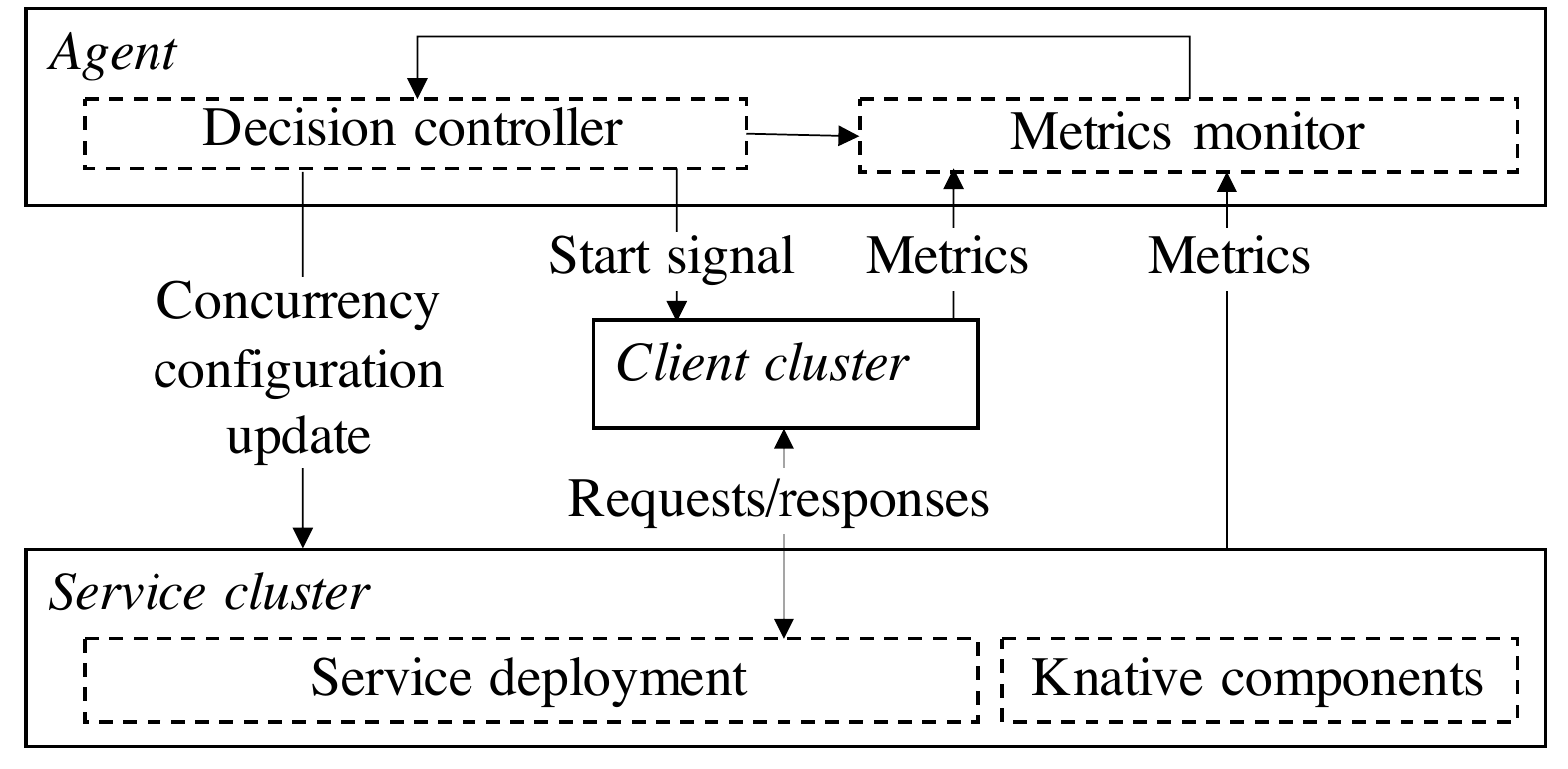}
\caption{Architectural setup including the information flow}
\label{fig:expsetup}
\end{figure}

The overall architecture of our experiment is illustrated in Fig. \ref{fig:expsetup}. 
To test the auto-scaling capabilities in an isolated environment, we set up two separate Kubernetes clusters, using IBM Cloud Kubernetes Service (IKS).
On the \textit{service cluster}, the sample service used for the experiments is deployed.
The cluster contains 9 nodes with 16 vCPU and 64 GB memory each, designed to provide sufficient capacity to host all Knative components and avoid performance limitations.
The \textit{client cluster} consists of one node with 16 vCPU and 64 GB memory responsible for sending requests to the service cluster to generate load.
The agent manages the activities on both clusters, including the configuration updates of the sample service based on collected metrics, and coordinates the process flow of the experiment, taking the role of an IKS user.

The Knative resources are installed on the service cluster (version v0.12), including the serving components explained in Section \ref{subsec:Knative}, which control the state of the deployed sample service and enable auto-scaling of additional pods on demand.
Using the trial-and-error method of RL in the second experiment, we update the concurrency configuration of the service in each iteration. 

To comprehensively test the auto-scaling capability, we activated the scale-to-zero functionality in the autoscaler's configmap, which requires a cold start in each iteration.
We further increased the replica number of ingress gateways, which handle load balancing, to bypass performance issues and to focus our studies exclusively on the auto-scaling functionalities.

\subsection{Workload}

Serverless computing is used for a variety of applications, accompanied by different resource requirements.
For example, the processing of video and image material or highly-parallel analytical workloads, such as MapReduce jobs, demand considerable memory and computing power. 
Other applications, such as chained API compositions or chatbots, tend to be less compute-intensive but may require longer execution or response time.

To investigate the concurrency impact of many different workloads, we generate a synthetic, stable workload profile simulating serverless applications.
We use Knative's example \emph{Autoscale-go} application for this purpose, which allows different parameters to be passed with the request to test incremental variations of the workload characteristics and thus emulate varying CPU- and memory- intensive workloads \cite{knativeautoscale-go}.
The three application parameters are \emph{bloat}, \emph{prime} and \emph{sleep}, wherein the first is used to specify the number of megabytes to be allocated and the second to calculate the prime numbers up to the given number, to create either memory- or compute-intensive loads.
The sleep parameter pauses the request for the corresponding number of milliseconds, as in applications with certain waiting times.

\subsection{Process Flow} \label{subsec:processflow}
\begin{figure}[t]
\centering
\includegraphics[width=8.5cm]{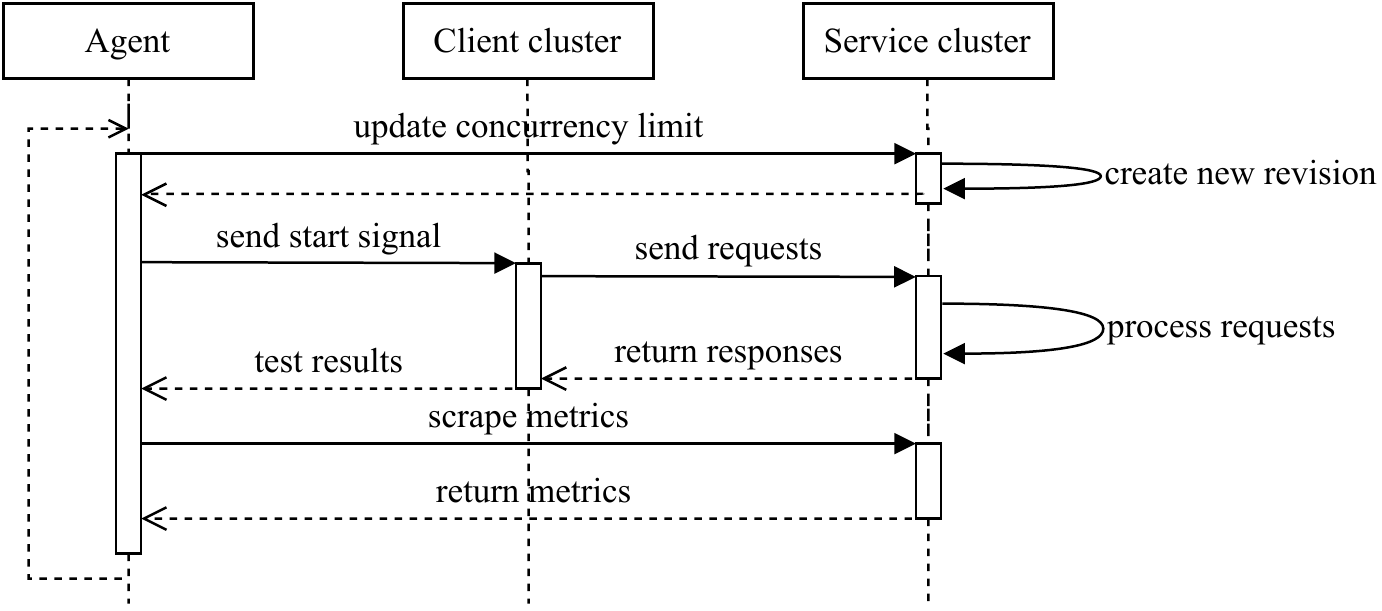}
\caption{Process flow}
\label{fig:processflow}
\end{figure}

The basic process flow of one iteration is illustrated in Fig. \ref{fig:processflow}.
In each iteration the agent sends a concurrency update to the service cluster, which accordingly creates a new revision with the respective concurrency limit.
When the service update is complete, the agent sends the start signal to the client cluster, which begins issuing parallel requests against the service cluster.
To simulate a large number of user requests at the same time, we use the HTTP load testing tool \emph{Vegeta}, which features sending HTTP requests at a constant rate.
In the experiment, 500 requests are sent simultaneously over a period of 30s to ensure sufficient demand for scaling and sufficient time to provide additional instances.
After the last response is received, Vegeta outputs a report of the test results, including information on latency distribution of requests, average throughput and success ratio of responses.
The performance measures are then stored by the agent.
Additionally, the agent crawls metrics from the Knative monitoring components, exposed via a Prometheus-based HTTP API within the cluster, to get further information about resource usage at cluster, node, pod and container level.
Using this data, the concurrency update is chosen to proceed to the next iteration.

%% file: sections/baselineexperiment.tex
 \label{sec:baseline}

To determine the implications of varying concurrency limits, we first conduct a baseline experiment comparing different workloads on their relative performance.

\subsection{Design}

As outlined in the previous section, we use the application parameters \emph{bloat}, \emph{prime} and \emph{sleep} to simulate varying workload characteristics. 
Starting with a no-operation workload where no parameters are passed, the memory allocation and CPU load were gradually increased for each new experiment.
The step size of the memory allocating parameter was aligned with the memory buckets commonly used for the standard pricing model of serverless platforms. To simulate compute-intensive and longer-lasting requests, different prime and sleep parameters were chosen correspondingly.
The detailed values are specified in \Cref{tab:concurrencyperformance}.

Per profile, we run performance tests for different concurrency levels according to the process flow described in \Cref{{fig:processflow}}. 
Theoretically, the concurrency limit can take all values between 0 and 1000.
To keep the experiments computationally feasible, we proceed in steps of 20, starting at a concurrency limit of 10 and ending at 310.
As stated in related literature, we focus on latency and throughput as key performance measures of serverless applications \cite{li2019understanding}. 
Average throughput is defined by requests per second (RPS), mean latency refers to the average time in seconds taken to return a response to a request.
To cover tail latency, we include the 95th percentile of latencies of all requests as an additional metric.
Furthermore, each test is repeated ten times to compensate for outliers or other fluctuations, before the concurrency is updated to the next limit.


\subsection{Results}

We structure the analysis of the baseline experiment results in three parts.
First, we examine the behavior of the individual workload profiles under different concurrency configurations.
Second, we focus on the relation of the target variables throughput and latency during the tests.
Finally, further metrics about resource utilization on container and pod level are analyzed.

As described above, we conducted the experiment for different combinations of the three parameters to simulate possible use cases. 
Table \ref{tab:concurrencyperformance} gives an overview of the outcomes with the concurrency limit that lead to the optimal test result in terms of one of the performance measures.
Due to the numerous uncontrollable factors that influence the performance of the cluster, each result forms a snapshot in time.
The respective workload configuration is described by the three columns on the left.
Taking all tests into account, the smallest possible concurrency of 10 is the most common configuration that resulted in the best performance across all three indicators.
Interestingly, this does not correspond to the default setting of the KPA where a target concurrency value of 100 is preferred \cite{knativekpaconfig}.
In particular, workloads that consume memory exclusively perform better with fewer parallel requests per pod instance, e.g tests \#\rom{2}, \#\rom{9}, \#\rom{16} and \#\rom{17}.
Similar observations are made for workloads with additional low CPU usage, i.e. lower \emph{prime} parameter, as in tests \#\rom{3} and \#\rom{10}.
Deviations can be observed when the requests pause for a certain time.
These workloads result in higher throughput and lower mean and tail latency when a higher concurrency is chosen, e.g tests \#\rom{7}, \#\rom{8} and \#\rom{14}.

\begin{table}[t]
\begin{threeparttable}
\caption{Concurrency Performance Tests }
\label{tab:concurrencyperformance}
\begin{centering}
\begin{tabular}{|c|c|c|c|c|c|c|}
\hline
\textbf{Test}&\multicolumn{3}{|c|}{\textbf{Workload Profile}}&\multicolumn{3}{|c|}{\textbf{Conc. Limit Yielding Best Perf.}} \\ 
\cline{2-7} 
\textbf{\#} & \textbf{\textit{bloat}}$^*$ & \textbf{\textit{prime}} &  \textbf{\textit{sleep}}$^*$ & \textbf{\textit{thrghpt}} & \textbf{\textit{mean lat.}} & \textbf{\textit{95th lat.}}\\
\hline
\rom{1} & - & - & - & 50 & 50 & 70 \\
\hline
\rom{2} & 128 & - & - & 30 & 30 & 30 \\
\hline
\rom{3} & 128 & 1000 & - & 10 & 10 & 10\\
\hline
\rom{4} & 128 & 10.000 & - & 30 & 30 & 10\\
\hline
\rom{5} & 128 & 100.000 & - & 10 & 10 & 10\\
\hline
\rom{6} &128 & 1000 & 1000 & 110 & 110 & 150 \\
\hline
\rom{7} & 128 & 10.000 & 1000 & 70 & 70 & 70 \\
\hline
\rom{8} & 128 & 100.000 & 1000 & 110 & 110 & 110 \\
\hline
\rom{9} & 256 & - & - & 10 & 10 & 10 \\
\hline
\rom{10} & 256 & 1000 & - & 10 & 10 & 10 \\
\hline
\rom{11} & 256 & 10.000 & - & 10 & 10 & 10 \\
\hline
\rom{12} & 256 & 100.000 & - & 10 & 10 & 10 \\
\hline
\rom{13} & 256 & 1000 & 1000 & 50 & 50 & 50 \\
\hline
\rom{14} & 256 & 10.000 & 1000 & 110 & 110  & 110 \\
\hline
\rom{15} & 256 & 100.000 & 1000 & 10 & 10  & 30 \\
\hline
\rom{16} &512 & - & - & 10 & 10 & 10\\
\hline
\rom{17} & 1024 & - & - & 30 & 30 & 30\\
\hline
\end{tabular}
\begin{tablenotes}
      \small
      \item $^*$ \textit{bloat} is defined in MB and \textit{sleep} in milliseconds.
    \end{tablenotes}
\label{tab1}
\end{centering}
\end{threeparttable}
\end{table}

\begin{figure}[t]
\centering
\includegraphics[width=7.5cm]{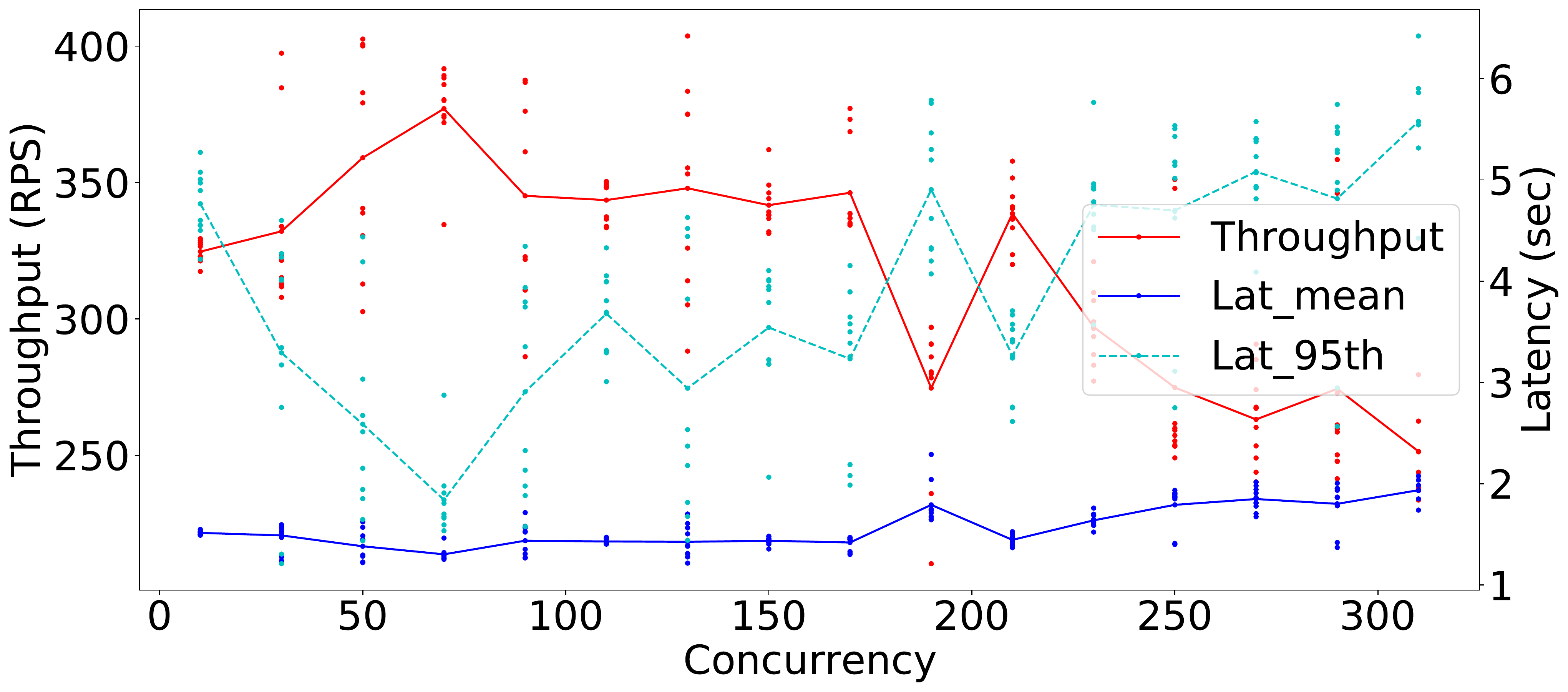}
\caption{Performance of workload test \#\rom{7}} 
\label{fig:thrpt_lat}
\end{figure}

Depending on the workload, the distance between the optimal configuration and the second best concurrency can be very small, which becomes more evident when analyzing a single test in detail.
Fig. \ref{fig:thrpt_lat} shows the result of test \#\rom{7}, which is examined representatively. 
Although the individual measurement points fluctuate, clear trends are identified in the average values.
A significant increase in throughput can be observed when the concurrency limit is raised to 70. 
This setting also yields the lowest value for mean latency, differing from the second-best value at concurrency 50 by only 80 milliseconds.
The distance becomes more critical when considering the tail latency of the $95^{th}$ percentile, where a request takes more than 740 milliseconds on average longer to receive a response when compared to the most effective configuration.
At a concurrency of 10, the difference amounts to almost 3 seconds, further underlining the performance variations caused by the different settings.
The greatest slowdown in tail latency in this test occurs at a concurrency of 310 with more than 3.7 seconds.

Besides, the overall performance decreases strongly when the concurrency limit exceeds a level of 210.
This tendency can be found across the majority of tests, indicating that due to the high simultaneous processing of many requests, only a limited amount of resources are available for a single request. 
Further observations show that with increasing memory utilization, i.e. the bloat parameter, performance tends to drop at lower concurrency limits.
In some cases, additionally, the success ratio strongly declines. 
For example in test \#\rom{16}, from a concurrency of 170 onwards, more than 10\% of the requests received non-successful responses. 
In test \#\rom{17} accordingly, this output can be observed from a concurrency of 90 onwards.

Focusing on the target metrics, the tests show that adjusting the concurrency limit to an appropriate setting can yield significant improvements in throughput and latency.
Furthermore, an inverse behavior of the measures can be observed within the tests.
For the previously considered test \#\rom{7}, the results indicate a significant negative correlation of throughput and mean latency of $-0.989$, and a similar correlation for throughput and $95^{th}$-percentile latency of $-0.916$.\footnote{For all statistical tests, Pearson correlation coefficient is used with a two-sided p-value for testing non-correlation and an alpha level of $.001$.}
This strong negative relationship between the metrics is found across all tests, with significant correlation coefficients ranging from $-0.995$ to $-0.748$.\footnote{Except for test \#\rom{1} and \#\rom{13} with significant correlations of throughput and tail latency of $-0.629$, and throughput and mean latency of $-0.677$.}
Subsequently, an improvement in throughput usually results in a lower and more favorable latency.
This finding implies that there is no need to make trade-offs between different target metrics when adjusting the concurrency. 
Instead, the problem can be reduced to one objective metric, representing the others.

%% file: sections/qlearningexperiment.tex
 \label{sec:qlearning}

The experiment described in previous section demonstrates the impact the concurrency configuration can have on performance.
Therefore, we evaluate the applicability of the model-free RL algorithm Q-learning in a second experiment to learn effective scaling policies by adjusting the concurrency limit during run time.

\subsection{Design}
The process flow is based on the procedure from section \ref{subsec:processflow}, extended with a more sophisticated logic of the agent.
Instead of incrementally increasing the concurrency, the agent uses knowledge of the system environment (states) to test different concurrency updates (actions) and evaluates them by receiving scores (reward).
In each iteration, the environment is defined by the current state, which, 
should provide a complete description of the system dynamics including all relevant information for optimal decision making.
Due to the large number of factors influencing performance, e.g. hidden cluster activities or network utilization, this is neither traceable nor computationally feasible in the used Q-learning algorithm.
Therefore, we break down our state space $S$ into three key features.
We define $S$ at time step $i$ as the combination of the state variables $s_i$ = $(conc_i, cpu_i, mem_i)$, where $conc_i$ depicts the concurrency limit, $cpu_i$ is the average CPU utilization per user-container and $mem_i$ is the average memory utilization per user-container.
The selection of the features is aligned with related research, with $conc_i$ as the equivalent of the number of VMs in VM auto-scaling approaches \cite{barrett2013applying, bitsakos2018derp}.
$cpu_i$ and $mem_i$ serve as a direct source of information about the resource utilization of a respective workload.
Since both CPU and memory utilization are continuous numbers, we discretize them into bins of equal size.
In each state $s_i \in S$, we define $A(s_i)$ as the set of valid actions, where $A$ is the set of all actions.
The agent can choose between decreasing, maintaining or increasing the concurrency limit by 20, i.e. $A$ = $\{-20,0,20\}$.
If the agent reaches the endpoints on the concurrency scale, i.e. the minimum or maximum concurrency, the action space in this state is reduced accordingly by the non-executable action.

After each iteration, the agent receives an immediate reward according to the performance achieved through the action.
In related literature, the reward is often based on the distance or ratio between the performance measure and a certain \textit{Service Level Agreement}, such as a throughput or response time target value \cite{horovitz2018efficient, dutreilh2011using}.
Since there is no target level to be achieved nor prior information about the performance given in our problem definition, we define an artificial reference value $ref\_value$ as the best value obtained to date.
Due to the permanent, albeit minor fluctuations in the measures, we propose a tolerance band around the reference value to avoid weighting minor non-relevant deviations.
Furthermore, the results from the preliminary study have shown a highly negative correlation between throughput and latency, i.e. higher throughput usually leads to lower and therefore better latency.
This relation in turn allows to focus exclusively on throughput ($thrghpt$) as one single objective.
The calculation of the reward $r$ in time step $i$ is as follows.
\[
  r_i = 
  \begin{cases}
     \frac{thrghpt_i}{ref\_{value}}
       & \begin{array}{r@{}}
            \text{if $thrghpt_i$} \leq \text{ref\_value} \cdot {0.95} \\ 
            \text{or $thrghpt_i$} \geq \text{ref\_value} \cdot {1.05}  
      \end{array}\\
       1 & \begin{array}{r@{}}
            \text{else}
      \end{array}\\
    \end{cases}
\]

Q-learning is initiated with the following parameters.
A learning rate $\alpha$ = 0.5 is chosen to balance newly acquired and existing information, a discount factor $\gamma$ = 0.9 to ensure that the agent strives for a long-term high return.
To encourage the exploration of actions at the beginning of training, we implement a decaying $\epsilon$-greedy policy starting at iteration 50 with $\epsilon$ = 1 and then slowly decrease over time by a decay factor of 0.995 per iteration.
The minimum exploration probability is set to $\epsilon_{min}$ = 0.1, to allow for the detection of possible changes in the system.
The knowledge the agent acquires is stored in a Q-table and updated each iteration.

To examine whether the model can effectively learn the concurrency values identified in section \ref{sec:baseline} as high throughput configurations, the results are analyzed representatively based on workload test \#\rom{7} and \#\rom{10}.
The former test showed high performance at a concurrency limit of 70, while the second reached the best test results at an edge concurrency of 10.

\subsection{Results}

\begin{figure}[t]
\centering
\includegraphics[width=8cm]{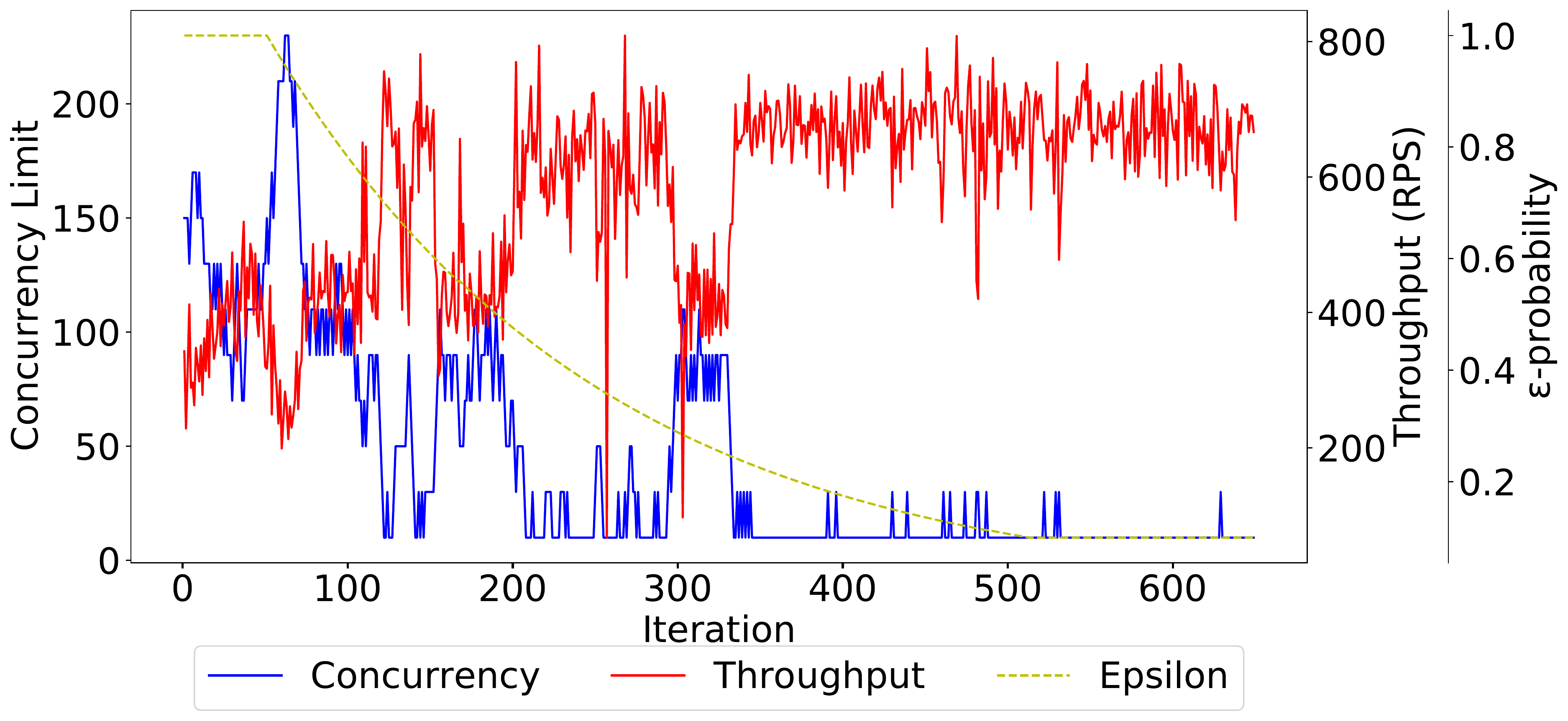}
\caption{Performance of Q-learning model, workload \#\rom{10}} 
\label{fig:qlearning1}
\end{figure}

\begin{figure}[t]
\centering
\includegraphics[width=8cm]{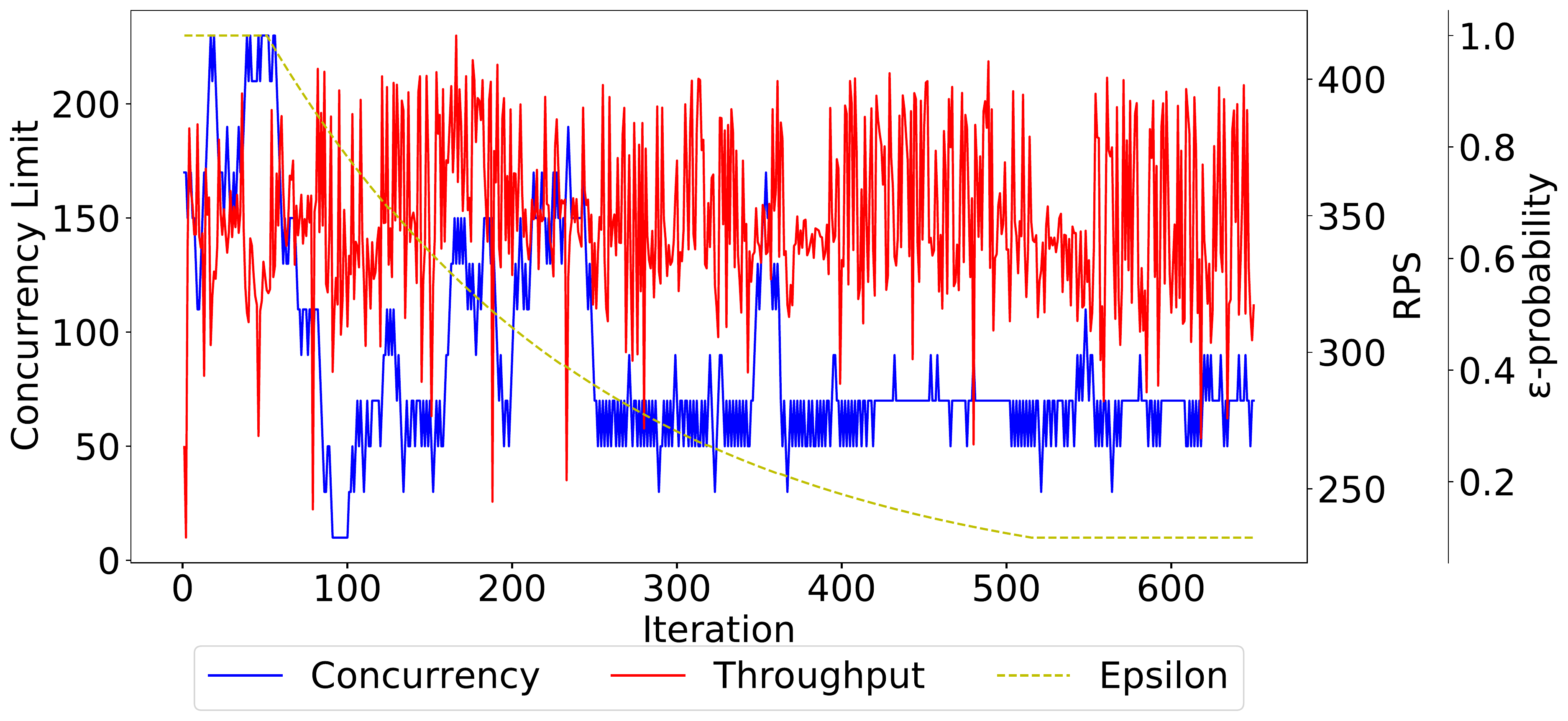}
\caption{Performance of Q-learning model, workload \#\rom{7}} 
\label{fig:qlearning2}
\end{figure}

First, we analyze the results to examine the suitability of the proposed Q-learning-based model to fine-tune the auto-scaling.
Second, we evaluate the performance of the approach in terms of throughput improvements compared to Knative's default auto-scaling configuration.

Based on workload profile \#\rom{10}, Fig. \ref{fig:qlearning1} shows in detail how the agent applies RL logic to incrementally change the concurrency and to adjust it as the training progressed. 
Beginning at concurrency 170, random exploration leads to a moderate decline of the concurrency limit in the first 30 iterations. 
This results in an improvement in throughput in this test, captured by the rewards and corresponding Q-values for each state-action combination.
The most effective scaling policy of 10 parallel requests per container is first reached in iteration 121. 
Nevertheless, due to the $\epsilon$-greedy strategy, exploratory actions are chosen, which might differ from the down-scaling decision and cause the agent to deviate from a good strategy.
As training progresses, a trend towards performance-enhancing concurrency configurations can be observed, indicating the agent is more likely to exploit the optimal decision rather than exploring.
After 330 iterations, the concurrency stabilizes at a limit of 10 parallel requests per container, implying the agent has learned the correct scaling policy, according to the results from section \ref{sec:baseline}.
Due to the minimum $\epsilon$ = 0.1, exploration still rarely occurs to ensure the agent can respond to changes in the environment.

A different learning process of the proposed Q-learning approach can be observed for workload \#\rom{7}, depicted in Fig. \ref{fig:qlearning2}.
The varying concurrency curve shows the initial strategy of the agent exploring first the higher state space before proceeding with lower concurrency limits.
After 250 iterations the exploitation phase outweighs and the concurrency gradually levels off.
In comparison to workload \#\rom{10}, where the algorithm's scaling policy converges to a single concurrency limit, the configuration here fluctuates, mainly between 50 and 70, and retains this pattern.
Further differences between the test results arise from the throughput metric, which shows strong fluctuations between 230 and 415 RPS across all iterations. 
The deviations, which also appear within one concurrency setting, considerably impair the agent's ability to evaluate suitable state-action-pairs via the reward function.
Nevertheless, the agent is able to narrow down the scaling range to a limited number of values at which it identified the best outcomes in terms of throughput, and which agrees with the result of the baseline experiment in Section \ref{sec:baseline}.

\begin{figure}[t]
\centering
\includegraphics[width=8cm]{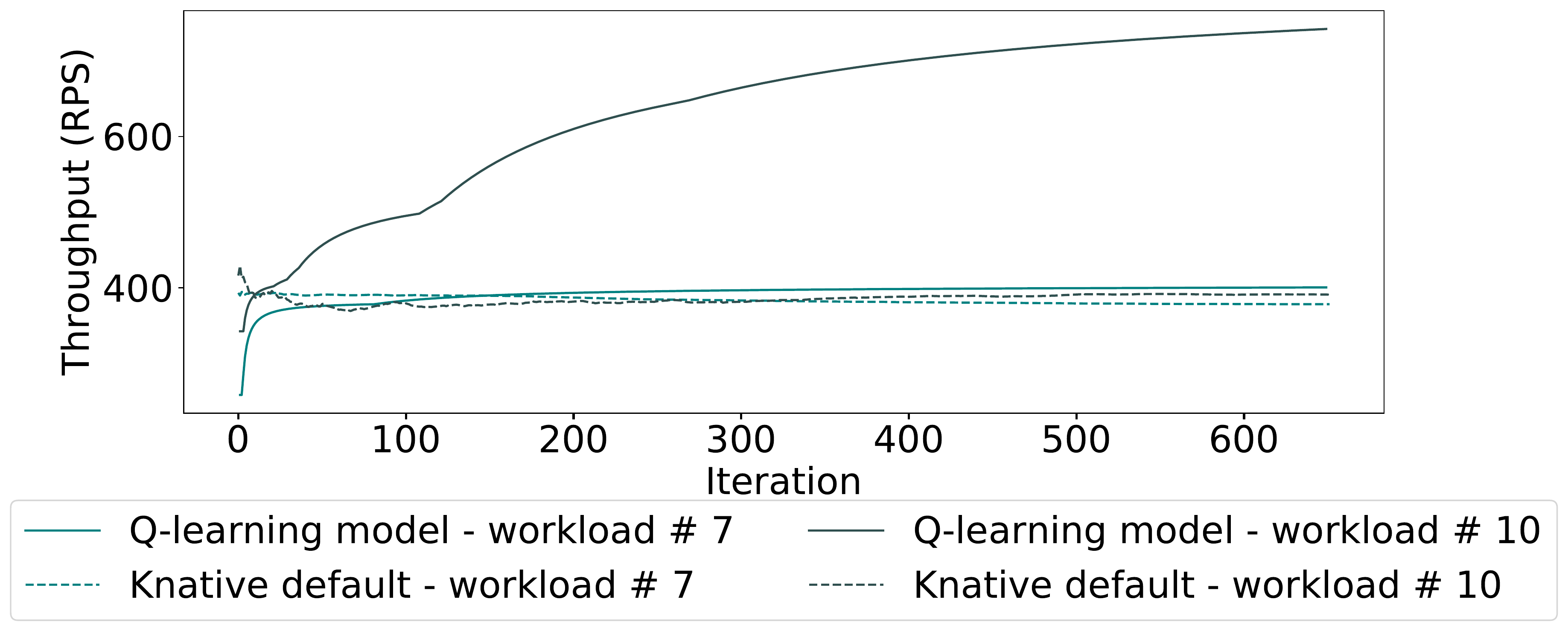}
\caption{Comparison of average throughput of the Q-learning model and the Knative default auto-scaling setting} 
\label{fig:qlearning_avgthrpt}
\end{figure}

To evaluate the proposed scaling policies, we benchmark the average performance of the Q-learning-based approach with the static default setting. 
For this purpose, the same experimental setup is used as in the Q-learning test, except for the auto-scaling configuration, where the original setting of a concurrency target of 100 is applied \cite{knativedocsautoscaling}.\footnote{Additionally, the container target percentage is set to 0.7 as in the default configmap.}
Fig. \ref{fig:qlearning_avgthrpt} depicts the average throughput up to the respective iteration of the Q-learning model and the default configuration for the two considered workloads. 
Both result in the Q-learning model outperforming the test based on Knative's standard settings.
Considering workload \#\rom{7} first, the model requires approximately 150 iterations until the average performance reaches default-level.
Subsequently, the throughput increases to an average of 400 RPS providing a minor advantage of 20 RPS compared to the standard system. 

A more significant enhancement shows workload \#\rom{10}.
While in the first 10 iteration the default settings alternate between 350 and 440 RPS, converging to an average of about 393 RPS, the performance of our model is initially lower.
However, with ongoing learning the average throughput improves and excels already from iteration 10 onwards. 
After 600 iterations, the presented Q-learning based model reaches an average throughput of 740 RPS, hence achieving more than 80\% of the performance of the default setting, which stabilizes at 390 RPS on average.

To summarize the results, the proposed model learned within finite time a scaling policy that outperforms the default Knative configuration in terms of throughput, proving the Q-learning-based approach is well-feasible to refine the auto-scaling mechanism.

%% file: sections/conclusion.tex
 \label{conclusion}

With the emergence of serverless frameworks, the ability of dynamic, real-time resource provisioning to meet varying demand has become a key area of interest and has led to the development of numerous scaling mechanisms.
Focusing on request-based scaling, we first investigated the impact a modification of the main scaling parameter, i.e. the number of concurrent requests per instance, may have on performance. 
The experiments showed deviations of up to multiple seconds in the average latency as well as significant differences in throughput values, thus indicating that the concurrency configuration can affect the performance depending on the specific workload.
To flexibly adjust the auto-scaling settings to specific requirements, we designed a RL model based on Q-learning and evaluated its applicability to learn effective scaling policies during runtime.
Based on different workloads, we showed that the proposed model can adapt the concurrency appropriately without prior knowledge within limited time and outperforms the average throughput compared to the default setting of Knative.

Given these results, the presented work offers valuable contributions to both the existing work in the field of serverless frameworks and the application of RL-based auto-scaling.
In addition to previous studies on scaling capabilities in serverless platforms, we provided a detailed analysis to reveal the performance implications of changes in the concurrency configuration.
Furthermore, we demonstrated, with our proposed model the applicability of Q-learning-based auto-scaling in the field of serverless applications.
Additionally, the findings can contribute to the ongoing development of the auto-scaling system of the Knative community project.

Nevertheless, we identified some limitations in the approach during the experiments.
First, the results from Section \ref{sec:baseline} are based on synthetic workloads simulated by one application with varying parameters, and thus cannot be interpreted as a universally valid conclusion on the effects of real-world applications.
Second, due to the focus on general applicability of Q-learning, the approach uses a rather simplistic reward function measuring exclusively the proximity to the reference value. 
Further refinement of the reward function may improve the efficiency of the proposed model.

While in this work a RL approach has been developed, which learns a certain scaling policy per workload mainly through testing different concurrency states, it remains to be analyzed to what extent the ratio of resource usage of individual components might impact the performance.
Thus, a comprehensive study could be conducted to determine the combination of utilization levels that might achieve the best possible performance across all workloads.
Consequently, the concurrency configuration could merely serve as a tool to bring the system into this particular state.